\documentclass{article}

\newcommand{\p}[1]{(\ref{#1})}

\newcommand{\cF}{{\cal F}}
\newcommand{\cE}{{\cal E}}

\newcommand{\cD}{{\cal D}}

\newcommand{\bZ}{{\overline Z}}

\newcommand{\bD}{{\overline D}}

\newcommand{\bQ}{{\overline Q}}
\newcommand{\bS}{{\overline S}}
\newcommand{\bW}{{\overline W}}
\newcommand{\bK}{{\overline K}}
\newcommand{\bJ}{{\overline J}{}}
\newcommand{\bLambda}{{\overline \Lambda}{}}

\newcommand{\bpsi}{{\bar\psi}{}}

\newcommand{\bnabla}{{\overline \nabla}}

\newcommand{\be}{\begin{equation}}
\newcommand{\ee}{\end{equation}}
\newcommand{\bea}{\begin{eqnarray}}
\newcommand{\eea}{\end{eqnarray}}

\newcommand{\ba}{\begin{array}} \newcommand{\ea}{\end{array}}

\def\im{{\rm i}}

\newcommand{\nn}{\nonumber}

\usepackage{amscd,amsmath,amssymb}

\def\theequation{\arabic{section}.\arabic{equation}}

\begin{document}
\thispagestyle{empty}
\vspace{2cm}
\begin{flushright}
\end{flushright}\vspace{2cm}
\begin{center}
{\Large\bf Component on shell actions of supersymmetric 3-brane\\[1cm]
 I. 3-brane in  D=6}
\end{center}
\vspace{1cm}

\begin{center}
{\large\bf S.~Bellucci${}^a$, N.~Kozyrev${}^b$, S.~Krivonos${}^{b}$ and A.~Sutulin${}^{a,b}$}
\end{center}

\begin{center}
${}^a$ {\it
INFN-Laboratori Nazionali di Frascati,
Via E. Fermi 40, 00044 Frascati, Italy} \vspace{0.2cm}

${}^b$ {\it
Bogoliubov  Laboratory of Theoretical Physics, JINR,
141980 Dubna, Russia} \vspace{0.2cm}

\end{center}
\vspace{2cm}

\begin{abstract}\noindent
In the present and accompanying papers we explicitly construct the on-shell supersymmetric component actions for
3-branes moving in D=6 and in D=8 within the nonlinear realizations framework.

In the first paper we apply our schema to construct the action of supersymmetric 3-brane in D=6. It turns out that all ingredients
entering the component action can be obtained almost algorithmically by using the nonlinear realizations approach. Within this approach, properly
adapted to the construction of on-shell component actions, we pay much attention to broken supersymmetry. Doing so, we were able to write
the action in terms of purely geometric objects (vielbeins and covariant derivatives of the physical bosonic components), covariant with respect to broken supersymmetry.
It turns out that all terms of the higher orders in the fermions, are hidden inside these covariant derivatives and vielbeins.
Moreover, the main part of the component action just mimics its bosonic cousin in which the ordinary space-time derivatives and the bosonic world volume are
replaced by their covariant supersymmetric analogs. The Wess-Zumino term in the action, which does not exist in the bosonic case, can be also easily constructed in terms of
reduced Cartan forms. Keeping the broken supersymmetry almost explicit, one may write the Ansatz for the component action, fully defined up to two constant parameters.
The role of the unbroken supersymmetry is just to fix these parameters.

\end{abstract}

\newpage
\setcounter{page}{1}
\setcounter{equation}{0}
\section{Introduction}
The usual treatment of the superbranes as extended objects arising as solitons of supersymmetric field theories, leads to the interpretation of the dynamics of branes fluctuations
as the field theories on the worldvolume of the superbrane. Thus, it seems to be important to have the explicit form of brane actions, preferably written in terms
of covariant objects. Unfortunately, the gauge fixing procedure in the standard Green-Schwartz-type formulation  is not unique. Moreover, the existing possibility
to choose different worldvolume fermions makes the task to write the full component superbrane actions with all fermions included rather complicated. The main problem in this way is to decide which type of covariant objects has to be taken to write the actions. Clearly, even if we would have at hands the proper superfield actions, after passing to the physical components
the beauty of the superspace formulation will disappear leaving us with a long tail of fermionic terms having no explicit geometric meaning.

The standard superspace description of the superbranes is based on the fact that usually the single branes preserve half of the target-space supersymmetries. Thus, it seems to be a quite
rational choice to introduce the superfields under unbroken supersymmetry, imposing by hands the invariance of the action under broken supersymmetry. One of the methods, heavily used
to describe the superbranes, is the method of nonlinear realization of spontaneously broken symmetries \cite{nlr1,nlr2}. In this approach, from the beginning we are dealing with
the theory in the static gauge in which the worldvolume fields, forming the proper supermultiplet of unbroken supersymmetry, correspond to the physical modes of the branes.
Within the nonlinear realization approach one may easily construct the objects, the Cartan forms and covariant derivatives, which are covariant with respect to unbroken and broken supersymmetries \cite{west1,west2,west3}. Moreover, by imposing the proper constraints on the Cartan forms, which are similar to those used in the superembeddings approach \cite{Dima1}, one may
find the irreducibility conditions on the superfields involved and the covariant superfield equations of motion (see e.g. \cite{rev1} and references therein). Nevertheless, one notices that the superfield action cannot be constructed within the nonlinear realization approach. The reason is simple - in all known cases the superfield Lagrangian  is not an invariant object.
Instead, it is shifted by the total space-time derivative under supersymmetry transformations. Thus, the superfield action cannot be constructed from the Cartan forms which are the natural ingredients of this approach. Just due to this reason, another approach in which the superfield Lagrangian is a component of the linear, with respect to both supersymmetries, supermultiplet has been elaborated \cite{bg2,bg3,rt,r,ik1,bik2,bik3,bik4}.

Coming back to the nonlinear realization method, one may wonder why the broken supersymmetry plays such a small role. Indeed, it is well known since the pioneering papers by D.~Volkov and
V.~Akulov \cite{VA} that the Goldstone fermions, accompanying the partial breaking of supersymmetry,  cannot enter the component action in an arbitrary way. They may appear in the action
either through the covariant derivatives or vielbeins, only. Therefore, one may try to shift the attention to the broken supersymmetry while dealing with the component on-shell actions.
Indeed, it was demonstrated in \cite{BIKS1} that with a suitable choice of the parametrization of the coset, the $\theta$ coordinates of the superspace together with the physical bosonic components do not transform under broken supersymmetry, while the physical fermions transform as the Goldstino of the Volkov-Akulov model. In such a situation, one should expect that
all fermionic terms in the components can be ``hidden'' inside the covariant derivatives and vielbeins, covariant with respect to broken supersymmetry. Therefore, one may
consider the reduced coset, which does not contain the $\theta$ coordinates at all. The corresponding reduced Cartan forms will be the building blocks for the component actions.
In the papers \cite{BKSu1,BIKS1,BIKS2,BIKS3} we explicitly demonstrated that this is indeed the case and the component on-shell actions for low dimensional branes ($D<4$ ) can be written through proper covariant derivatives of the bosonic fields and the fermionic vielbeins. In the present paper we analyze the more interesting cases with the 3 brane moving in $D=6$ and $D=8$. In the case of 3-branes the worldvolume effective actions  will be the actions for standard four-dimensional field theories, while the corresponding superfields will be the most interesting ones: the chiral superfields (in $D=6$) and hypermultiplet (in $D=8$).

This paper is divided into two parts. In this first part, in Section 2, we will find the superfield equations of motion and the transformation properties of the coordinates and superfields
within the standard method of nonlinear realization. Then, in Section 3, we will explicitly construct the component action of the 3-brane in $D=6$ and check its invariance with respect to broken and unbroken supersymmetries.

The second part of the paper will be devoted to the construction of the 3-brane action in $D=8$ with the hypermultiplet as the Goldstone superfield.

\section{Three-brane in $D=6$}
It is well known that the action of the $N=1$ supersymmetric three-brane in $D=6$ is just a minimal action for the
Goldstone chiral $N=1$ superfields accompanying the spontaneous breaking of $N=1, D=6$ supersymmetry down to $N=1, D=4$ one, or, in other words, the breaking
of $N=2$ supersymmetry to $N=1$ in four dimensions \cite{bg1,bg2,rt,r}.
In principle, this information is enough to construct the superfield action of the three-brane in terms of chiral $N=1$ superfields. The existence
of hidden $N=1$ supersymmetry completely fixes the action. However, passing to the on-shell component action almost completely destroys its very nice
superfields form. Moreover, it seems there is no way to combine the long tail of fermionic terms in some more readable form which would reflect the presence of higher symmetry in the theory. Our goal in this Section is to demonstrate that one may reformulate the on-shell component action in such a way that all
its ingredients will have the clear geometric properties. The basic tool for this is the method of nonlinear realizations \cite{nlr1,nlr2} adopted for this task in \cite{BKSu1}. Our procedure includes three steps: a) construction of the superfield equations of motion, b) checking the  bosonic action, c) construction of the full component action.

\subsection{Superfield equations of motion}
From the $d=4$ standpoint the $N=1, D=6$ supersymmetry algebra is a two central charges extended $N=2$  Poincar\'e superalgebra with the following basic relations
\be\label{superP}
\left\{ Q_{\alpha},  \bQ_{\dot\alpha}   \right\} =\left\{ S_{\alpha},  \bS_{\dot\alpha}   \right\} =2 \left( \sigma^A \right)_{\alpha\dot\alpha} P_A, \quad \left\{ Q_{\alpha},  S_{\beta}   \right\} =2 \epsilon_{\alpha\beta} Z,\; \left\{  \bQ_{\dot\alpha}, \bS_{\dot\beta} \right\} =2\epsilon_{\dot\alpha\dot\beta} \bZ.
\ee
As a reminder about its six-dimensional nature, the superalgebra \p{superP} possesses the $so(1,5)$ automorphism algebra. Again, from $d=4$ point of view, $so(1,5)$ algebra contains  the $d=4$ Lorentz algebra $(L_{AB})$, $u(1)$ subalgbera $(U)$ and the generators $K_A, \bK_A$ from the coset $SO(1,5)/SO(1,3)\times U(1)$. The full set of commutation relations can be found in the Appendix A.

Keeping the $d=4$ Lorentz and $U(1)$ symmetries linearly realized, we will choose the coset element as
\be\label{4dcoset}
g = e^{\im x^A P_A }e^{\theta^\alpha Q_{\alpha} + \bar\theta{}^{\dot\alpha}\bQ_{\dot\alpha}}  e^{\psi^\alpha S_{\alpha} + \bar\psi{}^{\dot\alpha}\bS_{\dot\alpha}} e^{\im(\varphi Z + \bar\varphi \bZ)} e^{\im( \Lambda^A K_A +\bLambda^A \bK_A)}.
\ee
Here, we associated the $N=1, d=4$ superspace coordinates $x^A, \theta^\alpha, \bar\theta{}^{\dot\alpha}$ with the generators $P_A, Q_\alpha, \bQ_{\dot\alpha}$ of unbroken $N=1$ supersymmetry. The remaining coset parameters are Goldstone superfields, $\psi^\alpha(x,\theta,\bar\theta)$,  $\bpsi^{\dot\alpha}(x,\theta,\bar\theta), \varphi(x,\theta,\bar\theta), \bar\varphi{}(x,\theta,\bar\theta)$.

The transformation properties of the coordinates and superfields with respect to all symmetries can be found by acting from the left on the coset element
$g$ \p{4dcoset} by the different elements of $N=2, d=4$ Poincar\'e supergroup. In particular, for the unbroken $(Q,\bQ)$ and broken $(S,\bS)$ supersymmetries we have
\begin{itemize}
\item Unbroken supersymmetry:
\be\label{susyQ}
\delta_Q x^A = \im \left( \epsilon^\alpha \bar\theta{}^{\dot\alpha}+ \bar\epsilon{}^{\dot\alpha} \theta^\alpha\right)\left( \sigma^A\right)_{\alpha\dot\alpha}, \quad
\delta_Q\theta^\alpha=\epsilon^\alpha, \; \delta_Q\bar\theta{}^{\dot\alpha}=\bar\epsilon{}^{\dot\alpha}.
\ee
\item Broken supersymmetry:
\be\label{susyS}
\delta_S x^A = \im \left( \varepsilon^{\alpha}\bpsi^{\dot\alpha} + \bar\varepsilon^{\dot\alpha}\psi^{\alpha}   \right)\left( \sigma^A  \right)_{\alpha\dot\alpha},\quad \delta_S \psi^\alpha = \varepsilon^\alpha, \; \delta_S\bpsi_{\dot\alpha} = \bar\varepsilon_{\dot\alpha},
\quad\delta_S \varphi = 2\im \varepsilon_{\alpha} \theta^{\alpha}, \ \delta_S \bar\varphi = 2\im \bar\varepsilon_{\dot\alpha} \bar\theta^{\dot\alpha}.
\ee
\end{itemize}
The local geometric properties of the system are specified by the Cartan forms. The purely technical calculations of these forms, semi-covariant derivatives and their algebra are summarized in the Appendix~A.

The next tasks, that we are going to perform within the superfield approach, include
\begin{itemize}
\item reduction of the number of independent superfields
\item imposing on the superfields the covariant irreducibility constraints
\item finding the covariant equations of motion.
\end{itemize}
As we already demonstrated in \cite{BKSu1}, all these tasks can be solved simultaneously by imposing the following constraints on the Cartan forms
\bea
&& \omega_Z = {\bar\omega}_Z=0,  \label{IH} \\
&& \left.  \omega_S\right|=\left . {\bar\omega}_S\right| =0, \label{dyn}
\eea
where $|$ means the $d\theta$ and $d\bar\theta$-projections of the forms. These constraints are similar to super-embedding conditions (see e.g. \cite{Dima1} and references therein).

The constraints \p{IH} are purely kinematical ones, and they result in the following equations
\bea\label{IH1}
&& \bnabla_{\dot\alpha} \varphi =0, \quad \nabla_\alpha \varphi = -2 \im \psi_\alpha ,  \qquad \nabla_A \varphi = 2\frac{(2+\lambda\bar\lambda) \lambda_A - \lambda^2 \bar\lambda_A }{(2+\lambda\bar\lambda)^2 -\lambda^2\bar\lambda^2}, \nn \\
&& \nabla_\alpha  \bar\varphi =0, \quad \bnabla_{\dot\alpha} \bar\varphi = -2 \im \bpsi_{\dot\alpha}, \qquad \nabla_A \bar\varphi = 2\frac{(2+\lambda\bar\lambda) \bar\lambda_A - \bar\lambda{}^2 \lambda_A }{(2+\lambda\bar\lambda)^2 -\lambda^2\bar\lambda^2} .
\eea
As we can see, these equations \p{IH1} allow us to express the superfields $\psi_\alpha, \bpsi_{\dot\alpha}$ and $\lambda_A ,\bar\lambda_A$ through the covariant
derivatives of $\varphi$ and $\bar\varphi$ (this is the so called Inverse Higgs phenomenon \cite{IH}). In addition, the superfields $\varphi$ and $\bar\varphi$ are subjected to
the covariant (anti)chirality conditions. Thus, the bosonic, covariantly (anti)chiral Goldstone superfields $\varphi$ and $\bar\varphi$ are the only essential superfields needed
for this case of partial breaking of the global supersymmetry.

The situation with the constraints \p{dyn} is more interesting. First, the $d\bar\theta$ ($d\theta$) projection of the form $\omega_S$ ($\bar\omega_S$) relates the spinor
derivative of the superfield $\psi$ ($\bar\psi$) and $x-$derivative of the superfield $\varphi$ ($\bar\varphi$)
\be\label{IH2}
J_{\alpha\dot\alpha} \equiv \bnabla_{\dot\alpha} \psi_\alpha = \partial_{\alpha\dot\alpha} \varphi + \ldots , \qquad
\bJ_{\alpha\dot\alpha} \equiv \nabla_\alpha \bpsi_{\dot\alpha} = \partial_{\alpha\dot\alpha} \bar\varphi + \ldots ,
\ee
where we explicitly write only the leading, linear in $\partial \varphi$ and $\partial \bar\varphi$ terms. At the same time,  the $d\theta$ ($d\bar\theta$) projection of the form $\omega_S$ ($\bar\omega_S$) gives the equations
\be\label{dyn1}
\nabla_{\alpha} \psi_\beta =0, \qquad \bnabla_{\dot\alpha} \bpsi_{\dot\beta}= 0.
\ee
To see that these equations are really equations of motion, note that from \p{dyn} and the algebra of covariant derivatives \p{alg_der} it follows that now
\be
\left\{ \nabla_\alpha , \nabla_\beta\right\} =0, \qquad \left\{ \bnabla_{\dot\alpha}, \bnabla_{\dot\beta}\right\} =0,
\ee
and, therefore, from \p{IH1} and \p{dyn1} we conclude that
\be\label{dyn2}
\nabla^\alpha \nabla_\alpha \varphi =0, \qquad \bnabla{}^{\dot\alpha} \bnabla_{\dot\alpha} \bar\varphi =0.
\ee
This is the covariant form of the superfield equations of motion, which is a proper covariantization of the free equations of motion.

In the next Sections we will construct the on-shell component action for our three-brane which gives the same component equations which follow from \p{dyn2}.

To close this Section, note that the explicit form of the equations \p{IH2}, which follows from the forms \p{dyn} is not very illuminating.  It is possible to get
more simple expressions for $J_{\alpha\dot\alpha}$ and $\bJ_{\alpha\dot\alpha}$ as follows. First of all, using the equations \p{dyn1}, one may rewrite the
anti-commutator $\{\nabla_{\alpha}, \bnabla_{\dot\alpha}\}$ \p{alg_der} as
\be\label{simpl1}
\{\nabla_{\alpha}, \bnabla_{\dot\alpha}\}= -2 \im \left(  \nabla_{\alpha\dot\alpha} +
 J_{\dot\alpha}^{\beta} \bJ_{\alpha}^{\dot\beta} \nabla_{\beta\dot\beta}\right).
\ee
Acting by this anti-commutator on $\varphi$ and $\bar\varphi$ and using \p{IH1} one may get\footnote{After passing to 4-vector notations, one should take into account, that
$\nabla_A \varphi$ is proportional to either $J_A$ or $\bJ_A$. Thus, the term with $\epsilon^{ABCD}J_B \bJ_C \nabla_D \varphi$ is zero.}
\bea\label{eqJ1}
&& J_A = \left( 1-  J_B \bJ^B \right)\nabla_A \varphi + \bJ^B \nabla_B\varphi\;J_A  + J^B\nabla_B \varphi \;\bJ_A, \nn \\
&& \bJ_A = \left( 1-  J_B \bJ^B \right)\nabla_A \bar\varphi + J^B \nabla_B\bar\varphi\;\bJ_A  + \bJ^B\nabla_B \bar\varphi \;J_A.
\eea
These equations can be easily solved for $\nabla_A \varphi, \nabla_A \bar\varphi$ as
\be\label{eqJ2}
\nabla_A \varphi = \frac{ J_A - J^2 \bJ_A}{1- J^2 \bJ{}^2}, \quad \nabla_A \bar\varphi = \frac{ \bJ_A - \bJ{}^2 J_A}{1-  J^2 \bJ{}^2},
\ee
where $J^2 = J^A J_A$ and $\bJ{}^2 = \bJ{}^A \bJ_A$. The expressions for $J_A, \bJ_A$ are more complicated to be
\be\label{eqJ2}
J_A = \nabla_A\varphi+ \frac{ 2 \left( \nabla\varphi\right)^2 \nabla_A \bar\varphi}{1- 2 \nabla\varphi\cdot \nabla\bar\varphi +
\sqrt{ \left( 1- 2 \nabla\varphi\cdot \nabla\bar\varphi\right)^2 - 4  \left( \nabla\varphi\right)^2 \left( \nabla\bar\varphi\right)^2}}, \quad
\bJ_A =\left( J_A\right)^\dagger.
\ee
\setcounter{equation}{0}
\section{Component action}
As we already noted in the Introduction, it is not clear how to construct the superfield action within the nonlinear realization approach.
For the supersymmetric 3-brane in $D=6$ the corresponding superfield actions were constructed within different frameworks in
\cite{bg2,rt}. Nevertheless, the component action, being constructed from the superfield one, is very complicated. It contains a lot of fermionic terms with
completely unclear geometric structure. Alternatively, as we demonstrated in \cite{BKSu1}, the component actions can be constructed within the nonlinear realization approach in such
way that the invariance with respect to broken supersymmetry becomes almost evident. The useful ingredients for this construction include the reduced Cartan forms and reduced
covariant derivatives, covariant with respect to broken supersymmetry only. The basic steps of our approach are
\begin{itemize}
\item construction of the bosonic action
\item covariantization of the bosonic action with respect to broken supersymmetry
\item construction of the Wess-Zumino terms
\item imposing the invariance with respect to unbroken supersymmetry.
\end{itemize}
Let us perform all these steps for the supersymmetric 3-brane in $D=6$.
\subsection{Bosonic action}
In principle, the bosonic equations of motion can be extracted from the superfield equations \p{dyn2}. But the calculations are rather involved. Instead, one can construct
the corresponding action directly, using the fact that such an action should possess invariance with respect to $D=6$ Poincar\'{e} symmetry spontaneously broken to
$d=4$. One of the key ingredients of such a construction is the bosonic limit of the Cartan forms \p{bos_forms} which explicitly reads
\bea\label{bos_forms1}
\left(\omega_P\right)^A_{bos} &=&  dx^B \left( \frac{1+y/2}{1-y/2} \right)_B^A  -2 \left( d\varphi \bar\lambda^B +d\bar\varphi \lambda^B \right)  \left( \frac{1}{1-y/2}  \right)_B^A, \nn\\
\left(\omega_Z\right)_{bos} &=& d\varphi + \left( d\varphi \bar\lambda^A + d\bar\varphi \lambda^A \right)\left( \frac{1}{1-y/2}  \right)_A^B \lambda_B  -  dx^A \left( \frac{1}{1-y/2}  \right)_A^B\lambda_B,  \\
\left(\bar\omega_Z\right)_{bos} &=& d\bar\varphi  + \left( d\varphi \bar\lambda^A + d\bar\varphi \lambda^A \right)\left( \frac{1}{1-y/2}  \right)_A^B \bar\lambda_B  - dx^A \left( \frac{1}{1-y/2}  \right)_A^B   \bar\lambda_B,\nn
\eea
where $ y_A^B = \lambda_A\bar\lambda{}^B+\bar\lambda_A \lambda^B$. Imposing now the same constraint \p{IH}
$$ \omega_Z =\bar\omega_Z=0,$$
we will get the bosonic analog of the relations \p{IH1}
\be\label{IHbos}
\partial_A \varphi = 2\frac{(2+\lambda\bar\lambda) \lambda_A - \lambda^2 \bar\lambda_A }{(2+\lambda\bar\lambda)^2 -\lambda^2\bar\lambda^2}, \quad
\partial_A \bar\varphi = 2\frac{(2+\lambda\bar\lambda) \bar\lambda_A - \bar\lambda{}^2 \lambda_A }{(2+\lambda\bar\lambda)^2 -\lambda^2\bar\lambda^2} .
\ee
Plugging these expressions in the form $\left(\omega_P \right)^A_{bos}$ \p{bos_forms1} and using the explicit expression for the matrix $\left(\frac{1}{1-y/2} \right)_A^B$
\be\label{Mmin1}
\left(\frac{1}{1-y/2} \right)_A^B = \delta_A^B +\frac{(2-\lambda\bar\lambda)\left( \lambda_A \bar\lambda^B +  \bar\lambda_A \lambda^B  \right) + \left(  \bar\lambda^2 \lambda_A \lambda^B + \lambda^2 \bar\lambda_A \bar\lambda^B     \right)}{(2-\lambda\bar\lambda)^2 -\lambda^2\bar\lambda^2},
\ee
one may obtain
\be\label{volumef1}
\left(\omega_P\right)^A_{bos} = dx^B\; e_B{}^A = dx^B \left(\delta_B^A - 2 \frac{(2+\lambda\bar\lambda)\left( \lambda_B \bar\lambda^A +  \bar\lambda_B \lambda^A  \right) - \left(  \bar\lambda^2 \lambda_B \lambda^A + \lambda^2 \bar\lambda_B \bar\lambda^A     \right)}{(2+\lambda\bar\lambda)^2 -\lambda^2\bar\lambda^2}\right).
\ee
Now, the unique invariant which can be constructed from the the forms $\left(\omega_P\right)^A_{bos}$ is the volume form which explicitly reads
\be\label{bos_action}
\epsilon_{ABCD} \left(\omega_P\right)^A_{bos}\wedge \left(\omega_P\right)^B_{bos}\wedge\left(\omega_P\right)^C_{bos}\wedge\left(\omega_P\right)^D_{bos} \sim
d^4x \det (e) = d^4x \frac{\left(2-\lambda\bar\lambda\right)^2 - \lambda^2 \bar\lambda{}^2}{\left(2+\lambda\bar\lambda\right)^2 - \lambda^2 \bar\lambda{}^2} .
\ee
Finally, the bosonic action, being rewritten in terms of $\partial\varphi$ and $\partial\bar\varphi$ acquires the form
\be\label{bos_action1}
S_{bos} = \int d^4 x \; \sqrt{\left(1 -2 \left( \partial \varphi \partial \bar\varphi\right)\right)^2  -4\left(\partial\varphi\partial\varphi\right)\left(\partial\bar\varphi\partial\bar\varphi\right)} .
\ee
This is the static gauge Nambu-Goto action for the 3-brane in $D=6$. One may explicitly check that the action \p{bos_action1} is invariant with respect
to $K_A,\bK_A$ transformations from the coset $SO(1,5)/SO(1,3)\times U(1)$ realized as
\be\label{trans1}
\delta x^A = 2\bar\alpha^A \varphi + 2 \alpha^A \bar\varphi, \qquad \delta \varphi =  \alpha_A x^A, \; \delta\bar\varphi = \bar\alpha{}_A x^A,
\ee
and therefore, it is invariant with respect to the whole $D=6$ Poincar\'{e} group.
\subsection{Covariantization  with respect to broken supersymmetry}
In contrast with the standard approach, in which the superfields with respect to unbroken $(Q, \bQ)$ supersymmetry play the main role and are the building blocks for
the superfield actions, in the component approach we are prefer to concentrate on the broken $(S, \bS)$ supersymmetry. Thus, the first task is to modify the bosonic action \p{bos_action1}
in such a way to achieve invariance with respect to broken supersymmetry. Due to the transformations laws \p{susyS}, the coordinates $x^A$ and the first components of the superfields $\varphi, \bar\varphi, \psi,\bar\psi$ transform under broken supersymmetry as follows
\be\label{susyS1}
\delta_S x^A = \im \left( \varepsilon^{\alpha}\bpsi^{\dot\alpha} + \bar\varepsilon^{\dot\alpha}\psi^{\alpha}   \right)\left( \sigma^A  \right)_{\alpha\dot\alpha},\quad \delta_S \psi^\alpha = \varepsilon^\alpha, \; \delta_S\bpsi^{\dot\alpha} = \bar\varepsilon^{\dot\alpha}, \quad\delta_S \varphi =0, \ \delta_S \bar\varphi = 0.
\ee
Thus, the volume $d^4x$ and the derivatives $\partial_A \varphi, \partial_A \bar\varphi$ are not the covariant objects. To find the proper objects, let us consider the reduced coset
element \p{4dcoset}
\be\label{4dcoset1}
g_{red} = e^{\im x^A P_A } e^{\psi^\alpha S_{\alpha} + \bar\psi{}^{\dot\alpha}\bS_{\dot\alpha}} e^{\im(\varphi Z + \bar\varphi \bZ)},
\ee
where the fields $\psi, \bar\psi, \varphi, \bar\varphi$ depend on the coordinates $x^A$ only. The corresponding reduced Cartan forms \p{formsA} read
\bea\label{formsA1}
&& \left( \omega_P\right)^A_{red} =\cE^A{}_B dx^B,\quad \cE^A{}_B \equiv \delta^A_B -\im \left(  \psi^\alpha \partial_B\bpsi^{\dot\alpha} + \bpsi^{\dot\alpha} \partial_B\psi^\alpha  \right) \left( \sigma^A \right)_{\alpha\dot\alpha} \nn \\
&&\left(\omega_Z\right)_{red} = d\varphi, \;\left(\bar\omega_Z\right)_{red} = d\bar\varphi,\;
(\omega_S)^\alpha_{red}  =  d\psi^\alpha,\; (\bar\omega_S)^{\dot\alpha}_{red}  =  d\bar\psi{}^{\dot\alpha}.
\eea
These forms are invariant with respect to transformations \p{susyS1}. Therefore, the covariant $x$-derivative will be
\be\label{cD}
\cD_A = \left(\cE^{-1}\right){}_A{}^B \partial_B,
\ee
while the invariant volume can be constructed from the forms $\left( \omega_P\right)^A_{red}$. Thus, the proper covariantization of the action \p{bos_action1}, having the right bosonic limit, will be
\be\label{actA}
S_1 = \int d^4 x \det (\cE)  \; \sqrt{\left(1 -2 \left( \cD \varphi \cD \bar\varphi\right)\right)^2  -4\left(\cD\varphi\cD\varphi\right)\left(\cD\bar\varphi\cD\bar\varphi\right)}.
\ee
The action $S_1$ \p{actA} reproduces the the fixed kinetic terms for bosons and fermions
\be
\left(S_1\right)_{lin} = \int d^4 x \left[ -\im \left(  \psi^\alpha \partial_{\alpha\dot\alpha}\bpsi^{\dot\alpha} + \bpsi^{\dot\alpha} \partial_{\alpha\dot\alpha}\psi^\alpha  \right)
+ 2 \partial^A \varphi \partial_A \bar\varphi \right].
\ee
This would be too strong to maintain unbroken supersymmetry. Therefore, we have to introduce one more, evidently invariant, action
\be\label{actB}
S_2= \alpha \int d^4x \; \det (\cE).
\ee
Thus, our anzatz for the invariant supersymmetric action of the 3-brane acquires the form
\be\label{actAB}
S= S_0+S_1+S_2 = \left( 1+ \alpha\right) \int d^4x - \int d^4 x \det (\cE) \left[\alpha  + \sqrt{\left(1 -2 \left( \cD \varphi \cD \bar\varphi\right)\right)^2  -4\left(\cD\varphi\cD\varphi\right)\left(\cD\bar\varphi\cD\bar\varphi\right)}\right],
\ee
where $\alpha$ is a constant that has to be defined, and we have added the trivial invariant action $S_0 = \int d^4x$ to have a proper limit
$$ S_{\varphi,\psi \rightarrow 0} =0.$$
\subsection{Wess-Zumino term}
The construction of the Wess-Zumino term, which is not strictly invariant, but which is shifted by a total derivative under broken supersymmetry \p{formsA1}, goes in a standard way \cite{luca}. First, one has to determine the close  five form $\Omega_5$, which is invariant under $d=4$ Lorentz and broken supersymmetry transformations \p{susyS1}. Moreover, in the present case
this form has to disappear in the bosonic limit, because our Ansatz for the action \p{actAB} already reproduces the proper bosonic action of the 3-brane \p{bos_action1}.
Such a form can be easily constructed in terms of the Cartan forms \p{formsA1}:
\be\label{wz1}
\Omega_5 = \omega_Z \wedge \bar\omega_Z \wedge \omega_S^\alpha \wedge \bar\omega_S^{\dot\alpha} \wedge \omega_P^{A} \; \left(\sigma_A\right)_{\alpha\dot\alpha}=
d\varphi \wedge d\bar\varphi \wedge d\psi^\alpha \wedge d \bar\psi{}^{\dot\alpha} \wedge \omega_P^{A}\; \left(\sigma_A\right)_{\alpha\dot\alpha}.
\ee
To see that $\Omega_5$ \p{wz1} is indeed closed, one should take into account that the exterior derivative of $\left(\omega_P\right)_{\alpha\dot\alpha}$ is given
by the expression
\be\label{add1}
d \left(\omega_P\right)_{\alpha\dot\alpha} \sim \left(\omega_S\right)_\alpha \wedge \left(\bar\omega_S\right)_{\dot\alpha} =
d\psi_\alpha \wedge d \bar\psi_{\dot\alpha},
\ee
and, therefore, $d\Omega_5=0$, because
$$ d\psi^\alpha \wedge d\psi_\alpha = d\bar\psi^{\dot\alpha} \wedge d\bar\psi_{\dot\alpha}=0.$$
Next, one has to write $\Omega_5$ as the exterior derivative of a 4-form $\Omega_4$:
\be\label{wz2}
\Omega_5 = d \Omega_4 \; \Rightarrow \; \Omega_4 = d\varphi \wedge d\bar\varphi \wedge \left( \psi^\alpha d \bar\psi^{\dot\alpha}+ \bar\psi^{\dot\alpha} d \psi^\alpha \right)  \wedge \omega_P^{A}\; \left(\sigma_A\right)_{\alpha\dot\alpha}.
\ee
Finally, the Wess-Zumino term is given by
\be\label{WZ}
S_{WZ} \sim \int \Omega_4 \sim \int d^4 x \det(\cE) \; \epsilon^{ABCD} \nabla_A \varphi \nabla_B \bar\varphi \left( \psi^\alpha \nabla_C \bar\psi^{\dot\alpha} +\bar\psi^{\dot\alpha}
\nabla_C\psi^\alpha\right)\; \left(\sigma_D\right)_{\alpha\dot\alpha}.
\ee
Before going further, let us note that the Wess-Zumino term \p{WZ} could be equivalently represented as
\be\label{WZ1}
S_{WZ} =\int d^4 x \det(\cE) \; \epsilon^{ABCD} \nabla_A \varphi \nabla_B \bar\varphi \left( \psi^\alpha \nabla_C \bar\psi^{\dot\alpha} +\bar\psi^{\dot\alpha}
\nabla_C\psi^\alpha\right) \left( \cE^{-1}\right)_D{}^E \; \left(\sigma_E\right)_{\alpha\dot\alpha}.
\ee
The proof is straightforward: if we substitute $\left( \cE^{-1}\right)_D{}^E$, given by
\be\label{Em1}
\left( \cE^{-1}\right)_D{}^E = \delta_D{}^E +\im \left(  \psi^\alpha \nabla_D\bpsi^{\dot\alpha} + \bpsi^{\dot\alpha} \nabla_D\psi^\alpha  \right) \left(\sigma^E\right)_{\alpha\dot\alpha}
\ee
in \p{WZ1}, the difference between the expressions \p{WZ} and \p{WZ1} will be
\be\label{proof1}
\sim
\int d^4 x \det(\cE) \; \epsilon^{ABCD} \nabla_A \varphi \nabla_B \bar\varphi \left( \psi^\alpha \nabla_C \bar\psi^{\dot\alpha} +\bar\psi^{\dot\alpha}
\nabla_C\psi^\alpha\right) \left( \psi_\alpha \nabla_D \bar\psi_{\dot\alpha} +\bar\psi_{\dot\alpha}\nabla_D\psi_\alpha\right)=0,
\ee
because
\be
\epsilon^{ABCD}\left( \psi^2 \nabla_C \bar\psi^{\dot\alpha} \nabla_D \bar\psi_{\dot\alpha}\right)=\epsilon^{ABCD}\left( \bar\psi^2 \nabla_C \psi^{\alpha} \nabla_D \psi_{\alpha}\right)=0
\ee
and
\be
\epsilon^{ABCD}\left( \nabla_C \psi^\alpha \psi_\alpha \; \nabla_D \bar\psi^{\dot\alpha} \bar\psi_{\dot\alpha}+
\nabla_D \psi^\alpha \psi_\alpha \; \nabla_C \bar\psi^{\dot\alpha} \bar\psi_{\dot\alpha}\right)=
\frac{1}{4} \epsilon^{ABCD}\left( \nabla_C(\psi^2) \nabla_D (\bar\psi^2) + \nabla_D(\psi^2) \nabla_C (\bar\psi^2)\right)=0.
\ee
As the last step, one has to prove that the Wess-Zomino term \p{WZ} or \p{WZ1} is invariant with respect to broken supersummetry transformations \p{susyS1}. For such a proof, the action \p{WZ1}
is more suitable, because its variation contains two pieces arising due to the shifts of  $\psi^\alpha$ and $\bar\psi^{\dot\alpha}$ explicitly presented in the integrand of $S_{WZ}$ \p{WZ1} and in $\left(\cE^{-1}\right)_A{}^B$ \p{Em1}:
\be\label{pr1}
\delta S_{WZ} = \delta_1 S_{WZ}+\delta_2 S_{WZ},
\ee
with
\be\label{pr2}
\delta_1 S_{WZ} =\int d^4 x \det(\cE) \; \epsilon^{ABCD} \nabla_A \varphi \nabla_B \bar\varphi \left( \varepsilon^\alpha \nabla_C \bar\psi^{\dot\alpha} +\bar\varepsilon^{\dot\alpha}
\nabla_C\psi^\alpha\right) \left( \cE^{-1}\right)_D{}^E \; \left(\sigma_E\right)_{\alpha\dot\alpha},
\ee
\be\label{pr3}
\delta_2 S_{WZ}=2 \im \int d^4 x \det(\cE) \; \epsilon^{ABCD} \nabla_A \varphi \nabla_B \bar\varphi \left( \psi^\alpha \nabla_C \bar\psi^{\dot\alpha} +\bar\psi^{\dot\alpha}
\nabla_C\psi^\alpha\right) \left(  \varepsilon_\alpha \nabla_D\bpsi_{\dot\alpha} + \bar\varepsilon_{\dot\alpha} \nabla_D\psi_\alpha  \right) .
\ee
The expressions \p{pr2} and \p{pr3} can be further simplified if we note that each covariant derivative $\nabla_A = (\cE^{-1})_A{}^B \partial_B$ contains $(\cE^{-1})$. Thus, these
four inverse veirbeins $(\cE^{-1})$, being converted with $\epsilon^{ABCD}$, give $\left( \det(\cE)\right)^{-1}$ which just cancels the factor $\det(\cE)$. Thus,
\be\label{pr4}
\delta_1 S_{WZ} \sim \int d^4 x  \; \epsilon^{ABCD} \partial_A \varphi \partial_B \bar\varphi \left( \varepsilon^\alpha \partial_C \bar\psi^{\dot\alpha} +\bar\varepsilon^{\dot\alpha}
\partial_C\psi^\alpha\right)  \; \left(\sigma_D\right)_{\alpha\dot\alpha},
\ee
\be\label{pr5}
\delta_2 S_{WZ} \sim \int d^4 x  \; \epsilon^{ABCD} \partial_A \varphi \partial_B \bar\varphi \left( \psi^\alpha \partial_C \bar\psi^{\dot\alpha} +\bar\psi^{\dot\alpha}
\partial_C\psi^\alpha\right) \left(  \varepsilon_\alpha \partial_D\bpsi_{\dot\alpha} + \bar\varepsilon_{\dot\alpha} \partial_D\psi_\alpha  \right) .
\ee
The integrand in \p{pr4} is obviously a full derivative, and therefore $\delta_1 S_{WZ}=0$. Next, the integrand in \p{pr5} can be rewritten as
\be\label{pr6}
\epsilon^{ABCD} \partial_A \varphi \partial_B \bar\varphi \left(\varepsilon_\alpha \psi^\alpha \partial_C\bpsi^{\dot\alpha} \partial_D \bpsi_{\dot\alpha}+
\bar\varepsilon_{\dot\alpha}\bpsi^{\dot\alpha} \partial_C\psi^\alpha\partial_D\psi_\alpha -\frac{1}{2}\left( \varepsilon_\alpha\partial_C \psi^\alpha \partial_D \bpsi^2 +
\bar\varepsilon_{\dot\alpha} \partial_C\bpsi^{\dot\alpha} \partial_D \psi^2\right)\right).
\ee
The first two terms in \p{pr6} are just zero, while the remaining two terms are full derivatives. Therefore,  $\delta_2 S_{WZ}=0$ and the Wess-Zumino term \p{WZ1} is invariant with respect to broken supersymmetry, as we expected.

To conclude, let us write the full Ansatz for the component action of 3-brane in $D=6$
\bea\label{actABC}
S& =& S_1+S_2 +S_{WZ} = \left( 1+ \alpha\right) \int d^4x - \int d^4 x \det (\cE) \left[\alpha  + \sqrt{\left(1 -2 \left( \cD \varphi \cD \bar\varphi\right)\right)^2  -4\left(\cD\varphi\cD\varphi\right)\left(\cD\bar\varphi\cD\bar\varphi\right)}\right] \nn \\
&& + \beta \int d^4 x \det(\cE) \; \epsilon^{ABCD} \cD_A \varphi \cD_B \bar\varphi \left( \psi^\alpha \cD_C \bar\psi^{\dot\alpha} +\bar\psi^{\dot\alpha}
\cD_C\psi^\alpha\right) \left( \cE^{-1}\right)_D{}^E \; \left(\sigma_E\right)_{\alpha\dot\alpha},
\eea
where $\alpha$ and $\beta$ are two constants that have to be defined by invariance with respect to unbroken supersymmetry.

Thus, after imposing broken supersymmetry, the component action \p{actABC} is fixed up to two constants $\alpha$ and $\beta$. No other terms in the action are admissible. The role of the unbroken supersymmetry is just to fix this constants.
\subsection{Unbroken supersymmetry}
In contrast with the standard superfield approach, the most technically complicated part of our approach is to maintain the unbroken supersymmetry, despite the fact
that all we need is to fix two constants in the action \p{actABC}. The first step in this task is to find the transformations of the needed ingredients in the action \p{actABC}
under unbroken supersymmetry \p{susyQ} (we will explicitly present here only the $\epsilon$-part of the transformations):
\bea\label{Qtransf}
&& \delta_Q \cD_A \varphi =2 \im\epsilon^\alpha \cD_A \psi_\alpha -2 \im \epsilon^\beta \cD_A \psi^\alpha \bJ_\beta{}^{\dot\alpha} \cD_{\alpha\dot\alpha} \varphi -
2 \im H^C \partial_C \cD_A \varphi, \nn \\
&& \delta_Q \cD_A \bar\varphi = -2 \im \epsilon^\beta \cD_A \psi^\alpha \bJ_\beta{}^{\dot\alpha} \cD_{\alpha\dot\alpha} \bar\varphi -
2 \im H^C \partial_C \cD_A \bar\varphi, \nn \\
&& \delta_Q \psi_\alpha = - 2 \im H^A \partial_A \psi_\alpha, \quad \delta_Q \bar\psi_{\dot\alpha} = -\epsilon^\alpha \bJ_{\alpha\dot\alpha} -2 \im H^A \partial_A \bpsi_{\dot\alpha}, \nn \\
&& \delta_Q \cD_A \psi_\alpha = -2 \im \epsilon^\beta \cD_A \psi^\gamma \bJ_\beta{}^{\dot\gamma} \cD_{\gamma\dot\gamma}\psi_\alpha -
2 \im H^C \partial_C \cD_A \psi_\alpha, \nn \\
&& \delta_Q \cD_A \bpsi_{\dot\alpha} = -\epsilon^\alpha \cD_A \bJ_{\alpha\dot\alpha}-2 \im \epsilon^\beta \cD_A \psi^\gamma \bJ_\beta{}^{\dot\gamma} \cD_{\gamma\dot\gamma}\bpsi_{\dot\alpha} -
2 \im H^C \partial_C \cD_A \bpsi_{\dot\alpha}, \nn \\
&& \delta_Q \left( \cE^{-1}\right)_A{}^B = 2 \im \cD_A H^B  -2 \im \epsilon^\beta \cD_A \psi^\gamma \bJ_\beta{}^{\dot\gamma} \left( \cE^{-1}\right)_D{}^B
\left(\sigma^D\right)_{\gamma\dot\gamma} - 2 \im H^C \partial_C \left( \cE^{-1}\right)_A{}^B,
\eea
where
\be\label{H}
H^{\alpha\dot\alpha} \equiv \epsilon^\beta \psi^\alpha \bJ_\beta{}^{\dot\alpha}.
\ee
As a consequence of \p{Qtransf} we will have
\be\label{Qtransfdet}
\delta_Q \det (\cE) = 2 \im \epsilon^\beta \cD_{\alpha\dot\alpha} \psi^\alpha \bJ_\beta{}^{\dot\alpha} \det (\cE) - 2 \im \partial_A \left( H^A \det (\cE)\right).
\ee
From \p{Qtransf} and \p{Qtransfdet} it immediately follows that in any action of the form
\be\label{testS}
S \sim \int d^4x \det(\cE) \cF\left[\cD \varphi, \cD\bar\varphi, \psi, \bpsi, \cD \psi, \cD\bpsi, \cE^{-1}\right]
\ee
with arbitrary function $\cF\left[\cD \varphi, \cD\bar\varphi, \psi, \bpsi, \cD \psi, \cD\bpsi, \cE^{-1}\right]$, the $H$-dependent terms will be converted into the full derivative
$-2 \im \partial_A \left( H^A \det (\cE)\; \cF \right)$. Therefore, one may ignore these parts of the transformations (except the $2 \im \cD_A H^B$ term in the transformations of $\cE{}^{-1}$)  , because our action \p{actABC} belongs to the class of actions in \p{testS}.

The parameter $\alpha$ can be defined if we will consider just the kinetic terms for $\varphi, \psi$ in the action \p{actABC}
\be\label{kin}
S_{kin}=\int d^4 x \left[ 2 \im (1+\alpha) \bpsi^{\dot\alpha} \partial_{\alpha\dot\alpha} \psi^\alpha+ 2 \partial^A \varphi \partial_A \bar\varphi \right].
\ee
The action \p{kin} has to be invariant with respect to linearized transformations \p{Qtransf}:
\be\label{Qlin}
\delta_Q \partial_A \varphi = 2 \im \epsilon^\alpha \partial_A \psi_\alpha, \; \delta_Q \partial_A \bar\varphi =0, \qquad \delta_Q \psi_\alpha=0, \;
\delta_Q \bpsi_{\dot\alpha} = - \epsilon^\alpha \partial_{\alpha\dot\alpha} \bar\varphi.
\ee
Varying the integrand in \p{kin} and integrating by parts, we will get
\be
\delta_Q S_{kin} = \int d^4x \left[ 2 \im (1+\alpha) \epsilon^\beta \bar\varphi\;  \partial_\beta{}^{\dot\alpha}\; \partial_{\alpha\dot\alpha}\psi^\alpha -
4 \im \epsilon^\alpha \bar\varphi\; \partial^A \partial_A \psi_{\alpha} \right].
\ee
Therefore, we have to fix
\be\label{alpha}
\alpha=1.
\ee
The fixing of the parameter $\beta$ is more involved. Ignoring the $H$-dependent terms in the transformation laws \p{Qtransf}, the variation of the
$S_1+S_2$ actions \p{actABC} with $\alpha=1$ under unbroken supersymmetry can be represented as
\be\label{part1}
\delta_Q (S_1+S_2) = \int d^4x \det (\cE)\;  \frac{4 \epsilon_\beta \cD_A \psi^\alpha}{1-J^2 \bJ^2} \left[ (1-J\cdot \bJ) \bJ_B \left(\sigma^{AB}\right)_\alpha^\beta+\bJ^A J_C \bJ_B  \left( \sigma^{CB}\right)_\alpha^\beta \right].
\ee
The full variation of the Wess-Zumino term is more complicated. But if we limit ourselves to the terms which are of the first order in fermions, this variation reads
\be\label{part2}
\delta_Q S_{WZ} \sim - 2\beta \int d^4 x  \frac{ \epsilon_\alpha \partial_A \psi^\beta}{1- J^2 \bJ^2} \left[ - J\cdot \bJ\; \bJ _B \left( \sigma^{AB} \right)_\alpha^\beta
+\bJ^2 J_B \left( \sigma^{AB} \right)_\alpha^\beta +\bJ^A J_B \bJ_C \left( \sigma^{BC} \right)_\alpha^\beta \right].
\ee
If we choose now $\beta=2$, then the variation of the full action acquires the form
\be\label{part3}
\delta_Q S \sim \int d^4 x   \frac{4 \epsilon_\beta \partial_A \psi^\alpha}{1-J^2 \bJ^2} \left( \bJ_B - \bJ^2 J_B\right) \left( \sigma^{AB} \right)_\alpha^\beta =
4 \int d^4 x\; \epsilon_\beta\; \partial_A \psi^\alpha \; \partial_B \bar\varphi \left( \sigma^{AB} \right)_\alpha^\beta.
\ee
Thus, the integrand in \p{part3} is a full derivative and, therefore, the action \p{actABC} is invariant in this approximation. The careful analysis of the terms with higher
order in the fermions shows that they also cancel out.
To conclude, let us write the full component action of 3-brane in $D=6$
\bea\label{actFIN}
S& =&  2 \int d^4x - \int d^4 x \det (\cE) \left[1  + \sqrt{\left(1 -2 \left( \cD \varphi \cD \bar\varphi\right)\right)^2  -4\left(\cD\varphi\cD\varphi\right)\left(\cD\bar\varphi\cD\bar\varphi\right)}\right] \nn \\
&& +2 \int d^4 x \det(\cE) \; \epsilon^{ABCD} \cD_A \varphi \cD_B \bar\varphi \left( \psi^\alpha \cD_C \bar\psi^{\dot\alpha} +\bar\psi^{\dot\alpha}
\cD_C\psi^\alpha\right) \left( \cE^{-1}\right)_D{}^E \; \left(\sigma_E\right)_{\alpha\dot\alpha},
\eea
This action is invariant with respect to both, broken and unbroken supersymmetries and, therefore, it is the supersymmetric 3-brane action in $D=6$.

\setcounter{equation}{0}
\section{Conclusion}
In this first part of our paper we have constructed the on-shell component action for $N=1, D=6$ supersymmetric 3-brane within the nonlinear realization approach.
We treated the worldvolume action of this 3-brane as the action of four-dimensional field theory which realized the partial breaking of $N=2, d=4$ global supersymmetry down to the $N=1, d=4$ one.
The $N=2, d=4$  Poincar\'{e} superalgebra we considered, contains two central charges which are just two translation generators from the $D=6$ point of view. These two translations, as well as half
of the $N=2, d=4$ supersymmetry, are supposed to be spontaneously broken. As the result of this spontaneous breakdown we have in the theory one complex bosonic and one fermionic Goldstone superfields in theory. The first
components of these superfields constitute the physical fields of the corresponding 3-brane. Within the superspace part of our consideration we have constructed the covariant irreducibility constraints on the
bosonic superfields and, in addition, we imposed the constrains which expressed the fermionic Goldstone superfields in terms of the bosonic ones. By using an analog of superembeddings constraints we were able
to find the covariant superfield equations of motion which have a proper bosonic equations of motion. This part is not new and has a close relation with the paper \cite{bg1} where similar
results were obtained. In the main part of this paper (Section 3) we have constructed the component action of this 3-brane. By shifting attention to the broken supersymmetry and considering the reduced coset, we
introduced the covariant derivatives and the fermionic vielbeins (covariant with respect to broken supersymmetry) which are the building blocks of the action. It turns out that all terms of higher orders in the fermions,
are hidden inside our covariant derivatives and vielbeins. Moreover, the main part of the component action just mimics its bosonic cousin in which the ordinary space-time derivatives and the bosonic world volume are
replaced by their covariant supersymmetric analogs. It is funny, that the Wess-Zumino term in the action, which does not exist in the bosonic case, can be also easily constructed in terms of
reduced Cartan forms. Keeping the broken supersymmetry almost explicit, one may write the Ansatz for the component action fully defined up to two constant parameters. The role of the unbroken
supersymmetry is to fix these parameters. Of course, the component action we have explicitly constructed in this paper, can be in principle obtained from the superfield action of the papers \cite{bg2,rt}.
Nevertheless, using the introduced covariant derivatives and fermionic vielbeins makes the action quite simple. Moreover, with our action one may, for example, perform the duality transformations, considered  in
\cite{r}, in full generality with all fermionic terms taken into account.

In the second part of this paper we will consider the supersymmetric 3-brane in $D=8$. From the $D=4$ point of view such a 3-brane corresponds to partial breaking of $N=4, d=4$ supersymmetry down to the $N=2, d=4$ one.
The corresponding Goldstone superfield, accompanying this breakdown of supersymmetry,  has to be a $N=2,d=4$ hypermultiplet, which makes such a system quite interesting.

\setcounter{equation}{0}
\section*{Acknowledgments}
The work of N.K. and S.K. was supported by RSCF grant 14-11-00598.

This work was partially supported by  the ERC Advanced Grant no. 226455 \textit{``Supersymmetry, Quantum Gravity and Gauge Fields''}~(\textit{SUPER\-FIELDS}).

\setcounter{equation}{0}
\def\theequation{A.\arabic{equation}}
\section*{Appendix A: Superalgebra, coset space, transformations and Cartan forms}
In this Appendix we collected some formulas describing the nonlinear realization
of $N=1, D=6$ Poincar\'{e} group in its coset over its $N=1, d=4$ subgroup.

In $d=4$ notation the $N=1, D=6$  Poincar\'e superalgebra is a two central charges extended $N=2$ super-Poincar\'{e} algebra containing the following set of generators:
\be\label{1}
\mbox{ N=2, d=4 SUSY }\quad \propto \quad \left\{ P_A, Q_\alpha,\bQ_{\dot\alpha}, S_\alpha,\bS_{\dot\alpha}, Z, \bZ, L_{AB}, K_{A}, \bK_{A}, U \right\}.
\ee
Here, $P_{A}, Z$  and $\bZ$ are $D=6$ translation generators,  $Q_{\alpha}, \bQ_{\dot\alpha}$ and $S_\alpha, \bS_{\dot\alpha}$ are the generators of super-translations, the generators $L_{AB}$ form $d=4$ Lorentz algebra $so(1,3)$, the generators $K_{A}$ and $\bK_{A}$ belong to the coset $SO(1,5)/SO(1,3)\times U(1)$, while $U$ span $u(1)$. The  commutation relations of $D=6$ Poincare algebra in this basis read
\bea\label{D6Poincare}
&&\left[L_{AB}, L_{CD}  \right]=\im \left( - \eta_{AC} L_{BD} + \eta_{BC} L_{AD} - \eta_{BD} L_{AC} +\eta_{AD} L_{BC}   \right),\; \left[ L_{AB}, P_{C}\right]= -\im \eta_{AC}P_{B} +\im \eta_{BC} P_{A};   \nn \\
&&\left[L_{AB}, K_{C}\right] = -\im \eta_{AC}K_{B} +\im \eta_{BC} K_{A}, \; \left[ L_{AB}, \bK_{C}\right] = -\im \eta_{AC}\bK_{B} +\im \eta_{BC} \bK_{A},  \nn  \\
&&\left[U, K_{A}\right] = K_{A}, \; \left[ U, \bK_{A} \right] = -\bK_{A}, \; \left[ U,Z  \right] =Z, \; \left[ U, \bZ  \right] = -\bZ;  \\
&& \left[K_{A}, \bZ\right] = \left[ \bK_{A}, Z  \right] =-2\im P_{A}, \left[ K_{A}, P_{B} \right]= -\im \eta_{AB} Z, \ \left[ \bK_{A}, P_{B} \right] = -\im \eta_{AB} \bZ; \nn \\
&& \left[ K_{A} , \bK_{B} \right] = 2\im L_{AB} -2\eta_{AB}U. \nn
\eea
Here, $\eta= \mbox{diag}(1,-1,-1,-1)$.

The four supercharges $Q_\alpha, \; \bQ_{\dot\alpha}, \; S_{\alpha}, \; \bS_{\dot\alpha}$ obey the following (anti)commutation relations
\bea\label{superD6Poincare}
&&\left\{ Q_{\alpha},  \bQ_{\dot\alpha}   \right\} =\left\{ S_{\alpha},  \bS_{\dot\alpha}   \right\} =2 \left( \sigma^A \right)_{\alpha\dot\alpha} P_A, \; \left\{ Q_{\alpha},  S_{\beta}   \right\} =2 \epsilon_{\alpha\beta} Z,\; \left\{  \bQ_{\dot\alpha}, \bS_{\dot\beta} \right\} =2\epsilon_{\dot\alpha\dot\beta} \bZ;\nn \\
&&\left[ L_{AB}, Q_\alpha   \right]  = -\frac{1}{2} \left( \sigma_{AB}  \right)_\alpha{}^\beta Q_\beta, \; \left[ L_{AB}, \bQ_{\dot\alpha}   \right]  = \frac{1}{2}  \bQ_{\dot\beta} \left(\tilde\sigma_{AB}  \right)^{\dot\beta}{}_{\dot\alpha},     \nn \\
&&\left[ L_{AB}, S_\alpha   \right]  = -\frac{1}{2} \left( \sigma_{AB}  \right)_\alpha{}^\beta S_\beta, \; \left[ L_{AB}, \bS_{\dot\alpha}   \right]  = \frac{1}{2}  \bS_{\dot\beta} \left(\tilde\sigma_{AB}  \right)^{\dot\beta}{}_{\dot\alpha};      \\
&&\left[ \bK_A, Q_\alpha    \right] = \im \left( \sigma_A  \right)_{\alpha\dot\alpha}\bS^{\dot\alpha}, \; \left[ \bK_A, S_\alpha    \right] = -\im \left( \sigma_A  \right)_{\alpha\dot\alpha}\bQ^{\dot\alpha},  \nn \\
&&\left[ K_A, \bQ_{\dot\alpha}    \right] = \im \left( \sigma_A  \right)_{\alpha\dot\alpha}S^{\alpha}, \; \left[ K_A, \bS_{\dot\alpha}    \right] = -\im \left( \sigma_A  \right)_{\alpha\dot\alpha}Q^{\alpha};  \nn \\
&&\left[ U, Q_\alpha\right] = \frac{1}{2}Q_\alpha, \; \left[ U, S_\alpha\right] = \frac{1}{2}S_\alpha, \; \left[ U, \bQ_{\dot\alpha}\right] = -\frac{1}{2}\bQ_{\dot\alpha}, \; \left[ U, \bS_{\dot\alpha}\right] = -\frac{1}{2}\bS_{\dot\alpha}. \nn
\eea
We define the coset element as follows
\be\label{4dcosetA}
g = e^{\im x^A P_A }e^{\theta^\alpha Q_{\alpha} + \bar\theta{}^{\dot\alpha}\bQ_{\dot\alpha}}  e^{\psi^\alpha S_{\alpha} + \bar\psi{}^{\dot\alpha}\bS_{\dot\alpha}} e^{\im(\varphi Z + \bar\varphi \bZ)} e^{\im( \Lambda^A K_A +\bLambda^A \bK_A)}.
\ee
Here, $\{ x^A, \theta^\alpha, \bar\theta{}^{\dot\alpha} \}$ are $N=1, d=4$ superspace coordinates, while the remaining coset parameters are Goldstone superfields.
The local geometric properties of the system are specified by the Cartan forms
\bea\label{cartanA}
g^{-1}dg &=& \im \left(\omega_P \right)^A P_A  + \im \omega_Z Z + \im \bar\omega_Z \bar Z +  \left( \omega_Q  \right)^\alpha Q_\alpha + \left( \bar\omega_Q  \right){}^{\dot\alpha} \bQ_{\dot\alpha} + \left( \omega_S  \right)^\alpha S_\alpha + \left( \bar\omega_S  \right){}^{\dot\alpha} \bS_{\dot\alpha}+ \nn \\
&&  \im  \left( \omega_K  \right)^A K_A +\im \left( \bar\omega_K  \right)^A \bK_A + \im \left( \omega_L  \right)^{AB} L_{AB}.
\eea
In what follows, we will need the explicit expressions of the following forms
\bea\label{formsA}
\left( \omega_P\right)^A &=& \triangle x^B\left(\cosh\sqrt{2Y}\right)_B^{A} - 2 \left( \triangle\varphi \bLambda^B + \triangle\bar\varphi \Lambda^B \right) \left( \frac{\sinh\sqrt{2Y}}{\sqrt{2Y}}  \right)_B^A , \nn \\
\omega_Z &=& \triangle\varphi  +  \left( \triangle\varphi \bLambda^A + \triangle\bar\varphi \Lambda^A \right) \left( \frac{\cosh\sqrt{2Y}-1}{Y}  \right)_A^B \Lambda_B  - \triangle x^A \left( \frac{\sinh\sqrt{2Y}}{\sqrt{2Y}}  \right)_A^B \Lambda_B , \nn\\
(\omega_Q)^\alpha &=& d\theta^\beta \left(\cosh \sqrt{W}\right)^\alpha_\beta +  d\bpsi^{\dot\alpha} \left( \frac{\sinh \sqrt{\bW}}{\sqrt{\bW}}  \right)^{\dot\beta }_{\dot\alpha} \Lambda_{\dot\beta}{}^\alpha,\nn \\
(\omega_S)^\alpha & = & d\psi^\beta \left(\cosh \sqrt{W}\right)_\beta^\alpha -  d\bar\theta^{\dot\alpha} \left( \frac{\sinh \sqrt{\bW}}{\sqrt{\bW}}  \right)^{\dot\beta }_{\dot\alpha} \Lambda_{\dot\beta}{}^\alpha,
\eea
where
\bea\label{forms_addA}
&&\triangle x^A = dx^A -\im \left( \theta^\alpha d\bar\theta^{\dot\alpha} + \bar\theta^{\dot\alpha} d\theta^\alpha + \psi^\alpha d\bpsi^{\dot\alpha} + \bpsi^{\dot\alpha} d\psi^\alpha  \right) \left( \sigma^A \right)_{\alpha\dot\alpha}, \nn \\
&& \triangle \varphi = d\varphi - 2\im \psi_{\alpha }d\theta^\alpha, \; \triangle \bar\varphi =d\bar\varphi - 2\im \bpsi_{\dot\alpha }d\bar\theta{}^{\dot\alpha}.
\eea
The matrix-valued functions $Y_A{}^B, W^\alpha{}_\beta, \bW^{\dot\alpha}{}_{\dot\beta}$ are defined as follows
\be\label{mat_functions_A}
Y_A{}^B = \Lambda_A \bLambda^B + \bLambda_A \Lambda^B, \quad
W^\alpha{}_\beta = \Lambda^{\alpha\dot\alpha}\bLambda_{\beta\dot\alpha}, \; \bW^{\dot\alpha}{}_{\dot\beta} = \bLambda^{\alpha\dot\alpha}\Lambda_{\alpha\dot\beta}.
\ee
For the bosonic forms $\left( \omega_P\right)^A, \omega_Z, \bar\omega_Z$ we find useful to perform the changing of the variables
\bea\label{stereogr}
\lambda^{A} = \left(\frac{\tanh{\sqrt{\frac{Y}{2}}}}{\sqrt{\frac{Y}{2}}} \right)^A_B \Lambda^B, \quad \bar\lambda^{A} = \left(\frac{\tanh{\sqrt{\frac{Y}{2}}}}{\sqrt{\frac{Y}{2}}} \right)^A_B \bLambda^B .
\eea
This is just a stereographic projection. Introducing now the new matrix $y_A^B = \lambda_A \bar\lambda^B + \bar\lambda_A \lambda^B$, and noting that
\bea\label{ymatrix}
Y_A^B = 2 \left( \mathrm{arctanh}^2\sqrt{\frac{y}{2}}   \right)_A^B,
\eea
one may rewrite the bosonic forms in \p{formsA} as
\bea\label{bos_forms}
\left(\omega_P\right)^A &=& \triangle x^B \left( \frac{1+y/2}{1-y/2} \right)_B^A  -2 \left( \triangle\varphi \bar\lambda^B +\triangle\bar\varphi \lambda^B \right)  \left( \frac{1}{1-y/2}  \right)_B^A, \nn\\
\omega_Z &=& \triangle\varphi + \left( \triangle\varphi \bar\lambda^A + \triangle\bar\varphi \lambda^A \right)\left( \frac{1}{1-y/2}  \right)_A^B \lambda_B  -  \triangle x^A \left( \frac{1}{1-y/2}  \right)_A^B\lambda_B,  \\
\bar\omega_Z &=& \triangle\bar\varphi  + \left( \triangle\varphi \bar\lambda^A + \triangle\bar\varphi \lambda^A \right)\left( \frac{1}{1-y/2}  \right)_A^B \bar\lambda_B  - \triangle x^A \left( \frac{1}{1-y/2}  \right)_A^B   \bar\lambda_B.\nn
\eea

Keeping in mind, that the  quantities $\triangle x^A, d\theta^\alpha$ and $d{\bar\theta}{}^{\dot\alpha}$ are invariant with respect to both supersymmetries \p{susyQ}, \p{susyS}, one may define the semi-covariant derivatives $\nabla_A, \nabla_{\alpha}, \bnabla_{\dot\alpha}$ as
\be\label{diff}
d \cF  = \left( dx^A \frac{\partial}{\partial x^A}+ d\theta^\alpha \frac{\partial}{\partial \theta^\alpha}+d\bar{\theta}^{\dot\alpha} \frac{\partial}{\partial \bar{\theta}^{\dot\alpha}} \right) \cF = \left( \triangle x^A \nabla_A +d\theta^\alpha \nabla_\alpha + d{\bar\theta}{}^{\dot\alpha} \bnabla_{\dot\alpha}\right) \cF,
\ee
and, therefore,
\bea\label{4dsusy5}
&&\nabla_A = (E^{-1})_A{}^B\; \partial_B, \ E_A{}^B = \delta_A{}^B -\im \left( \psi^\alpha \partial_A \bpsi^{\dot\alpha} +  \bpsi^{\dot\alpha} \partial_A \psi^\alpha\right)\left( \sigma^B \right)_{\alpha\dot\alpha},  \\
&& \nabla_\beta = D_\beta -\im \left( \psi^\alpha D_\beta \bpsi^{\dot\alpha} +  \bpsi^{\dot\alpha} D_\beta \psi^\alpha\right)\left( \sigma^B \right)_{\alpha\dot\alpha}\nabla_B = D_\beta -\im \left( \psi^\alpha \nabla_\beta \bpsi^{\dot\alpha} +  \bpsi^{\dot\alpha} \nabla_\beta \psi^\alpha\right)\left( \sigma^B \right)_{\alpha\dot\alpha}\partial_B,
\nn \\
&& \bnabla_{\dot\beta} =\bD_{\dot\beta} -\im \left( \psi^\alpha \bD_{\dot\beta} \bpsi^{\dot\alpha} +  \bpsi^{\dot\alpha} \bD_{\dot\beta} \psi^\alpha\right)\left( \sigma^B \right)_{\alpha\dot\alpha}\nabla_B = \bD_{\dot\beta} -\im \left( \psi^\alpha \bnabla_{\dot\beta} \bpsi^{\dot\alpha} +  \bpsi^{\dot\alpha} \bnabla_{\dot\beta} \psi^\alpha\right)\left( \sigma^B \right)_{\alpha\dot\alpha}\partial_B. \nn
\eea
These derivatives satisfy the following (anti)commutation relations
\bea\label{alg_der}
&&\left\{ \nabla_\alpha, \nabla_\beta \right\} = -2\im \left( \nabla_\alpha \psi^\gamma \nabla_\beta \bpsi^{\dot\gamma} + \nabla_\beta \psi^\gamma \nabla_\alpha \bpsi^{\dot\gamma} \right) \left( \sigma^C \right)_{\gamma \dot\gamma} \nabla_C, \nn \\
&& \left[  \nabla_A , \nabla_B \right] = 2\im \left( \nabla_A \psi^\gamma \nabla_B \bpsi^{\dot\gamma} - \nabla_B \psi^\gamma \nabla_A \bpsi^{\dot\gamma} \right) \left( \sigma^C \right)_{\gamma \dot\gamma} \nabla_C, \\
&& \left\{ \nabla_\alpha, \bnabla_{\dot\beta} \right\}= -2\im \left( \delta^\gamma_\alpha \delta^{\dot\gamma}_{\dot\beta}+ \nabla_\alpha \psi^\gamma \bnabla_{\dot\beta} \bpsi^{\dot\gamma} + \bnabla_{\dot\beta} \psi^\gamma \nabla_\alpha \bpsi^{\dot\gamma} \right)\left( \sigma^C \right)_{\gamma \dot\gamma} \nabla_C, \nn \\
&& \left[ \nabla_A, \nabla_\alpha \right] = -2\im \left( \nabla_A \psi^\gamma \nabla_\alpha \bpsi^{\dot\gamma} +  \nabla_\alpha \psi^\gamma \nabla_A \bpsi^{\dot\gamma}  \right) \left( \sigma^C \right)_{\gamma\dot\gamma} \nabla_C. \nn
\eea
Here, $D_\alpha, \bD_{\dot\alpha}$ are flat covariant derivatives  obeying the relations
\be
\left\{ D_\alpha, \bD_{\dot\alpha}\right\} = -2 \im \left(\sigma^A\right)_{\alpha\dot\alpha} \partial_A, \quad
\left\{ D_\alpha, D_{\beta}\right\}=\left\{ \bD_{\dot\alpha}, \bD_{\dot\beta}\right\}=0.
\ee

Finally, we will define the transition to the vectors as
\be\label{vector_A}
V^A =\frac{1}{2}  V^{\alpha\dot\alpha} \left( \sigma^A\right)_{\alpha\dot\alpha}.
\ee
We used the following definitions of the $\sigma$-matrices:
\be\label{4dsigma}
\left(\sigma^A \right)_{\alpha\dot\alpha} = \left( 1, \vec \sigma  \right), \; \left(\tilde\sigma^A \right)^{\alpha\dot\alpha}=\epsilon^{\alpha\beta}\epsilon^{\dot\alpha\dot\beta} \left(\sigma^A \right)_{\beta\dot\beta}  = \left( 1, -\vec \sigma  \right) .
\ee
($\vec \sigma$ is the ordinary set of three-dimensional Pauli matrices).
Indices of these matrices and spinors are raised and lowered by $\epsilon_{\alpha\beta}, \; \epsilon_{\dot\alpha\dot\beta}$ with properties
\be\label{4dspeps}
\epsilon_{\alpha \gamma} \epsilon^{\gamma \beta} =  \delta_{\alpha}^{\beta}, \; \epsilon_{\dot\alpha \dot\gamma} \epsilon^{\dot\gamma \dot\beta} =  \delta_{\dot\alpha}^{\dot\beta}, \quad
\epsilon_{12} = \epsilon_{\dot 1 \dot 2} = 1.
\ee
Therefore,
\be\label{app2}
(\sigma_A)_{\alpha \dot{\alpha}} (\tilde{\sigma}_B)^{\dot{\alpha} \alpha}
= {\rm Tr} (\sigma_A \tilde{\sigma}_B) = 2 \eta_{AB}, \quad (\sigma^A)_{\alpha \dot{\alpha}} (\tilde{\sigma}_A)^{\dot{\beta} \beta} = 2\, \delta_{\alpha}^{\beta} \delta_{\dot\alpha}^{\dot\beta},
\ee
and
\bea\label{app5}
&&\left(\sigma^A \right)_{\alpha \dot\alpha} \left(\tilde{\sigma}^B\right)^{\dot\alpha \beta} = \eta^{AB} \delta^{\beta}_{\alpha} - \im \left(\sigma^{AB}\right)_{\alpha}{}^{\beta}, \quad \left(\tilde{\sigma}^A\right)^{\dot\beta \alpha} \left(\sigma^B\right)_{\alpha \dot\alpha} = \eta^{AB} \delta^{\dot\beta}_{\dot\alpha}
- \im \left(\tilde{\sigma}^{AB}\right)^{\dot\beta}{}_{\dot\alpha}, \nn \\
&&
\left( \sigma_A \tilde{\sigma}_B \sigma_C\right)_{\alpha\dot\alpha} =
  \eta_{AB} \left (\sigma_C \right)_{\alpha \dot\alpha}+ \eta_{BC} \left (\sigma_A \right)_{\alpha \dot\alpha}- \eta_{AC} \left(\sigma_B \right)_{\alpha \dot\alpha}+ \im \epsilon_{ABCD} \left (\sigma^D \right)_{\alpha \dot\alpha},\nn\\
&&
\left( \tilde{\sigma}_A \sigma_B \tilde{\sigma}_C \right)^{\dot\alpha \alpha} =  \eta_{AB} \left(\tilde{\sigma}_C\right)^{\dot\alpha \alpha} + \eta_{BC}  \left(\tilde{\sigma}_A \right)^{\dot\alpha \alpha}- \eta_{AC}  \left(\tilde{\sigma}_B \right)^{\dot\alpha \alpha}- \im \epsilon_{ABCD}  \left(\tilde{\sigma}^D \right)^{\dot\alpha \alpha}\,,
\eea
where two-indices matrices are defined as
\bea\label{4dspeps2}
\left( \sigma^{AB} \right)_\alpha{}^\beta = \frac{\im}{2} \left[ \left( \sigma^A  \right)_{\alpha\dot\alpha} \left( \tilde\sigma^B  \right)^{\dot\alpha\beta}- \left( \sigma^B  \right)_{\alpha\dot\alpha} \left( \tilde\sigma^A  \right)^{\dot\alpha\beta}   \right], \nn \\
\left( \tilde\sigma{}^{AB} \right)^{\dot\beta}{}_{\dot\alpha} = \frac{\im}{2} \left[ \left( \tilde\sigma^A  \right)^{\dot\beta\alpha}\left( \sigma^B  \right)_{\alpha\dot\alpha} -  \left( \tilde\sigma^B  \right)^{\dot\beta\alpha}\left( \sigma^A  \right)_{\alpha\dot\alpha}   \right].
\eea
Note, that the matrices $\sigma^{AB}\; (\tilde\sigma{}^{AB})$ are self-dual (anti-self-dual), respectively
\be\label{app7}
 \left(\sigma^{AB}\right)_{\alpha}{}^{\beta} = -\frac{\im}{2} \epsilon^{ABCD} \left(\sigma_{CD}\right)_{\alpha}{}^{\beta}, \;
\; \left(\tilde{\sigma}^{AB}\right)^{\dot\beta}{}_{\dot\alpha} =  \frac{\im}{2}\, \epsilon^{ABCD} \left(\tilde{\sigma}_{CD}\right)^{\dot\beta}{}_{\dot\alpha}.
\ee
Finally, we define the $d=4$ volume form in a standard manner as
\be\label{volume}
d^4 x \equiv \epsilon_{ABCD}dx^A \wedge dx^B \wedge dx^C \wedge dx^D \quad \Rightarrow \quad
dx^A \wedge dx^B \wedge dx^C \wedge dx^D = -\frac{1}{24} \epsilon^{ABCD} d^4x,
\ee
with
\be
\epsilon_{0123} =- \epsilon^{0123} =1.
\ee

\end{document}